\documentclass{article}
\usepackage{spconf,amsmath,amssymb,graphicx,hyperref}
\usepackage{booktabs}
\usepackage{dsfont}
\graphicspath{{figures/}}

\begin{document}
\title{ModalFidelity: Routing Modalities for Deepfake Detection on a Budget}

\name{Oguzhan Baser$^{\dagger *}$ \qquad Kaan Kale$^{\ddagger}$ \qquad
      Sriram Vishwanath$^{\ddagger}$ \qquad Sandeep Chinchali$^{\dagger}$
\thanks{This work has been submitted to the IEEE for possible publication. Copyright may be transferred without notice, after which this version may no longer be accessible. $^{*}$Corresponding author. Email: oguzhanbaser@utexas.edu.
This work was supported in part by the National Science Foundation under Grant 2148186.
Any opinions, findings, and conclusions or recommendations expressed in this material are those of
the authors and do not necessarily reflect the views of the National Science Foundation.}}

\address{$^{\dagger}$The University of Texas at Austin, Austin, TX, USA \\
         $^{\ddagger}$Georgia Institute of Technology, Atlanta, GA, USA}

\ninept
\maketitle

\begin{abstract}
Deepfakes no longer need to fake a whole video. Generators that read the transcript now alter only the few seconds in which a video's meaning turns, so a forgery hides in a small, unknown fraction of the video. Yet detectors still read every one-second window of both the audio and image streams, spending nearly all of their compute where nothing was altered. We observe that deciding where to look is far cheaper than looking. We present ModalFidelity, a lightweight router that previews each window and decides, before any forensic detector runs, which stream is worth reading, under a hard compute budget it can never exceed. On AV-Deepfake1M, reading at most a fifth of the windows, it is more accurate than gating after the detectors at 15.9$\times$ less compute, and retains over 96\% of the accuracy of an oracle that knows where every forgery lies.
\end{abstract}

\begin{keywords}
multimodal deepfake, modality selection, budgeted inference, adaptive computation, temporal forgery
\end{keywords}

\section{Introduction}

Modern generative models synthesize media that people can no longer reliably tell from a recording \cite{baser2024securespectra}. A single open-source stack clones a voice from seconds of audio, animates a photorealistic face, and aligns lip motion to the forged speech \cite{jia2018sv2tts, siarohin2019fomm, prajwal2020wav2lip}, so a forgery is rarely confined to one channel. What has changed recently is not fidelity but \textit{selectivity}. Reasoning-capable generators now read a transcript, locate the few words on which meaning turns (e.g., a name, a dosage, a negation), and spend their compute only there \cite{cai2024avdeepfake1m, baser2025phonemefake}. This inverts the economics of an attack. Faking a two-minute call once meant synthesizing all two minutes; now it takes only the two seconds that change its meaning. Such attacks have already caused real financial losses: in one reported case, a finance worker transferred \$25M after a video call in which every other participant was synthetic \cite{chen2024deepfakecfo}. Such \textbf{targeted forgeries} are cheap enough to mount during a live exchange, and because the altered span is short and semantically chosen, it is at once harder to find and more damaging when missed. Detection inherits the harder half of this bargain: it must locate the few altered seconds. Streams arrive continuously and are acted upon once, so a verifier that runs every detector over every second of every stream cannot keep pace. It spends nearly all of that effort on windows where nothing was altered.

\begin{figure}[t]
  \centering
  \includegraphics[width=0.9\linewidth]{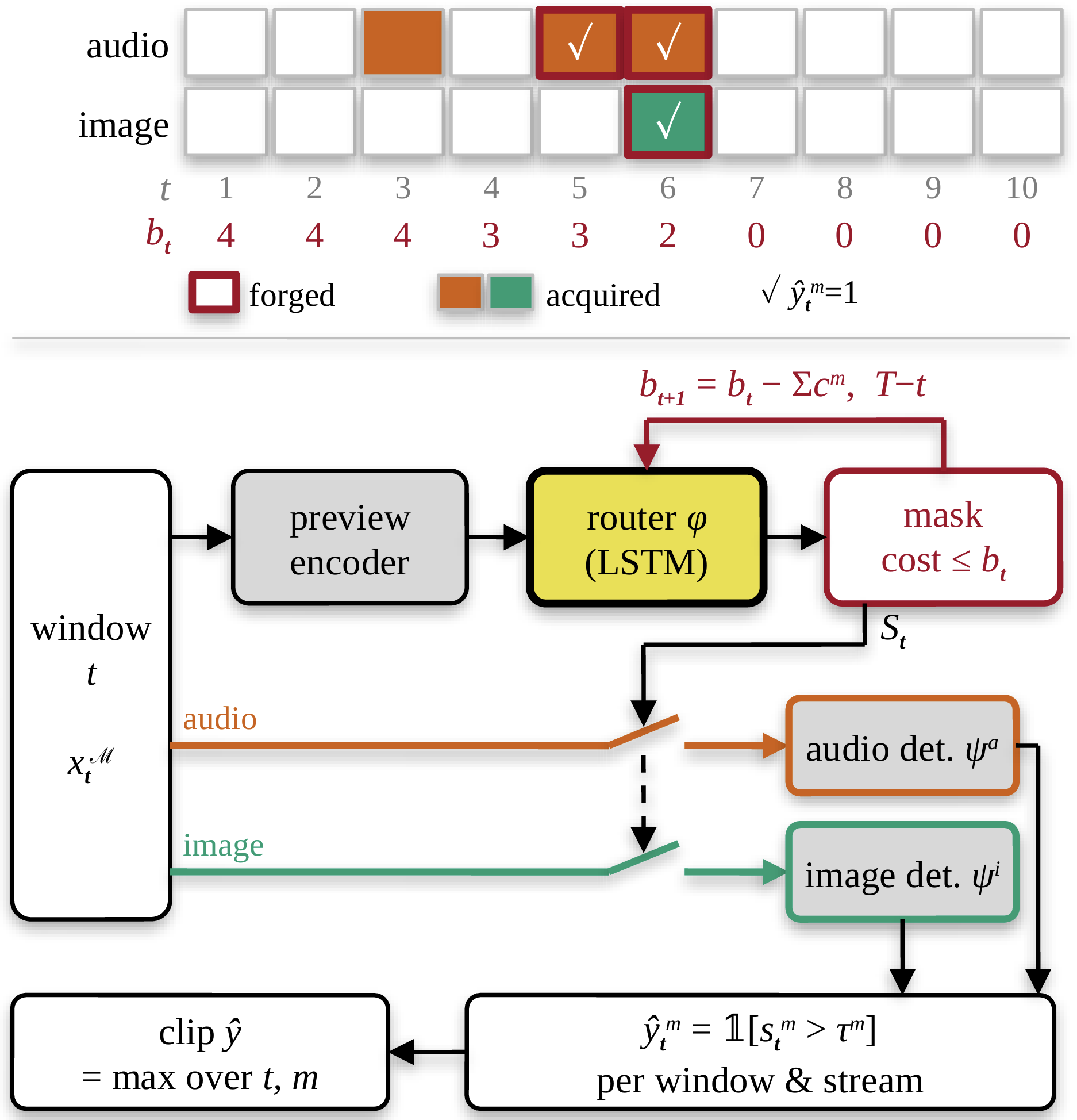}
  \caption{\textbf{How can our router cut cost without cutting accuracy?} \textit{Bottom:} a cheap preview of each window $t$ feeds the router $\varphi$ (yellow, the only trained part), which picks the streams $S_t$ to acquire; a mask removes any action costing more than the remaining budget $b_t$ (red). Only acquired streams reach the frozen detectors $\psi^a,\psi^i$ (grey), whose per-window, per-stream predictions $\hat y_t^m$ are max-pooled into the clip prediction $\hat y$. \textit{Top:} one video with budget $B{=}4$; the router saves its budget for the forged windows (red outline) and flags all three forged streams.}
  \label{fig:arch}
\end{figure}

Despite the prevalence and threat of such multimodal attacks, a \textit{central open problem} is \textbf{detecting and localizing forgeries that occupy a small, unknown fraction of the stream}. Existing detectors fall into two families, and neither decides which windows and streams are worth analyzing. First, \textit{single-modality, whole-clip classifiers} return one prediction per sample \cite{jung2022aasist, rossler2019ffpp}. They are blind to manipulation in the channel they do not read. An audio model cannot see a swapped face. Because the prediction is global, they cannot report \textit{when} the forgery occurred, which is precisely what a forensic account requires. Second, \textit{early- and late-fusion multimodal detectors} consume every stream and are correspondingly more accurate \cite{baltrusaitis2019survey, zhou2021joint, mittal2020emotions}, but they evaluate every encoder on every window. Their cost grows with duration times the number of modalities, and nearly all of it is spent where nothing was altered. Averaging over a long genuine remainder even dilutes the very evidence a short forgery leaves behind. What is missing is a detector that jointly provides \textit{(i)} selective acquisition under an explicit compute budget, \textit{(ii)} localization in time \textbf{and} modality, and \textit{(iii)} a cost the verifier fixes in advance, a small fraction of the cost of reading every window.

We \textbf{observe} that the utility of information is \textit{dynamic along both axes} of a multimodal stream. Evidence is sparse in time, since a manipulated window is the exception rather than the rule, and it is asymmetric across modalities, since a targeted edit usually leaves its trace in one stream and not the other. Uniform effort is therefore the wrong default: it pays the same price for a window that decides the outcome as for one that cannot possibly inform it. Our \textbf{key insight} is that \textit{deciding where to look is far cheaper than looking}. A preview encoder reading a handful of frames and a spectrogram costs a small fraction of a forensic detector, so a policy placed \textbf{before} acquisition can afford to inspect everything cheaply in order to spend expensively almost nowhere. We therefore ask: \textit{how can we verify only where it matters to expose tailored deepfake forgeries at a fraction of the cost?} Thus, we propose \textbf{ModalFidelity}, a lightweight router that decides window by window which modality to acquire and when to stop, under an explicit compute budget (Fig.~\ref{fig:arch}).\\
\textbf{\underline{Related work}} addresses parts of this problem, but not all of it. \textit{Adaptive computation}, such as early exit and token pruning, saves effort inside a model but assumes the input has already been acquired \cite{teerapittayanon2016branchynet, rao2021dynamicvit}. \textit{Multimodal deepfake detectors} do localize manipulations \cite{cai2022batfd}, yet run every encoder on every window. \textit{Mixture-of-experts (MoE) routing} selects among experts \cite{shazeer2017moe}, but a router placed after them can only redistribute a cost already paid.\\

\noindent In light of prior work, our \textbf{contributions are threefold:}
\begin{itemize}
    \item We design a routing model that decides, window by window, which stream is worth reading under a hard compute budget it can never exceed, and is more accurate than gating after the detectors at \textbf{15.9}$\times$ less compute.
    \item We systematically study where forgery evidence lies across time and streams and how accuracy grows with the budget, finding that our router retains over \textbf{96\%} of the accuracy of an oracle that knows where every forgery lies.
    \item We release our code\footnote{\href{https://github.com/UTAustin-SwarmLab/modal-fidelity}{\texttt{github.com/UTAustin-SwarmLab/modal-fidelity}}\label{code}} including our router model, its training pipeline, and a benchmark that pins down \textit{where} gating pays off in a detector stack: before the detectors, not after them.
\end{itemize}

\section{Methodology}
\label{sec:method}

Consider a video call in which the chief financial officer orders an urgent \$25M transfer. A \textbf{provider} streams audio and video to a \textbf{verifier}, which must release or hold the transfer \textit{before the call ends}. An \textbf{adversary} need not fake the whole meeting: altering the seconds that carry the account number, or the face that approves it, is enough. A prediction delivered afterwards arrives after the money, and one reporting only \textit{that} something was forged leaves the verifier nothing to act on, since it cannot tell which second was faked.

\vspace{2pt}\noindent\textbf{\underline{Problem formulation:}} We treat a stream as a sequence of $T$ fixed-length windows indexed by $t\in\{1,\dots,T\}$, each carrying $M$ co-occurring modalities $\mathcal{M}$; here audio and image, $\mathcal{M}=\{a,i\}$, so $M{=}2$. Let $x_t^m$ denote the content of modality $m$ in window $t$ and $x_t^{\mathcal{M}}$ the window as a whole. Each modality has a frozen forensic detector $\psi^m$ that returns a score $s_t^m=\psi^m(x_t^m)$ at a cost of $c^m$ units. Supervision consists of a clip label $y\in\{0,1\}$, predicted by $\hat y$, and per-window, per-stream labels $y_t^m\in\{0,1\}$ marking whether stream $m$ was manipulated at window $t$. ModalFidelity's router $\varphi(\cdot;\theta_\varphi)$ (Fig.~\ref{fig:arch}) reads a cheap preview of the windows seen so far, together with the remaining budget $b_t$ and horizon $T{-}t$, and commits to an acquisition set
$S_t=\varphi\big(x_{1:t}^{\mathcal{M}},\,b_t,\,T{-}t;\,\theta_\varphi\big)\subseteq\mathcal{M}$, possibly empty, where $b_t=B-\sum_{t'<t}\sum_{m\in S_{t'}}c^m$ is what remains of a budget $B$, set as a fraction $\rho$ of the windows, $B\approx\rho T$.

\begin{figure}[t]
  \centering
  \includegraphics[width=\linewidth]{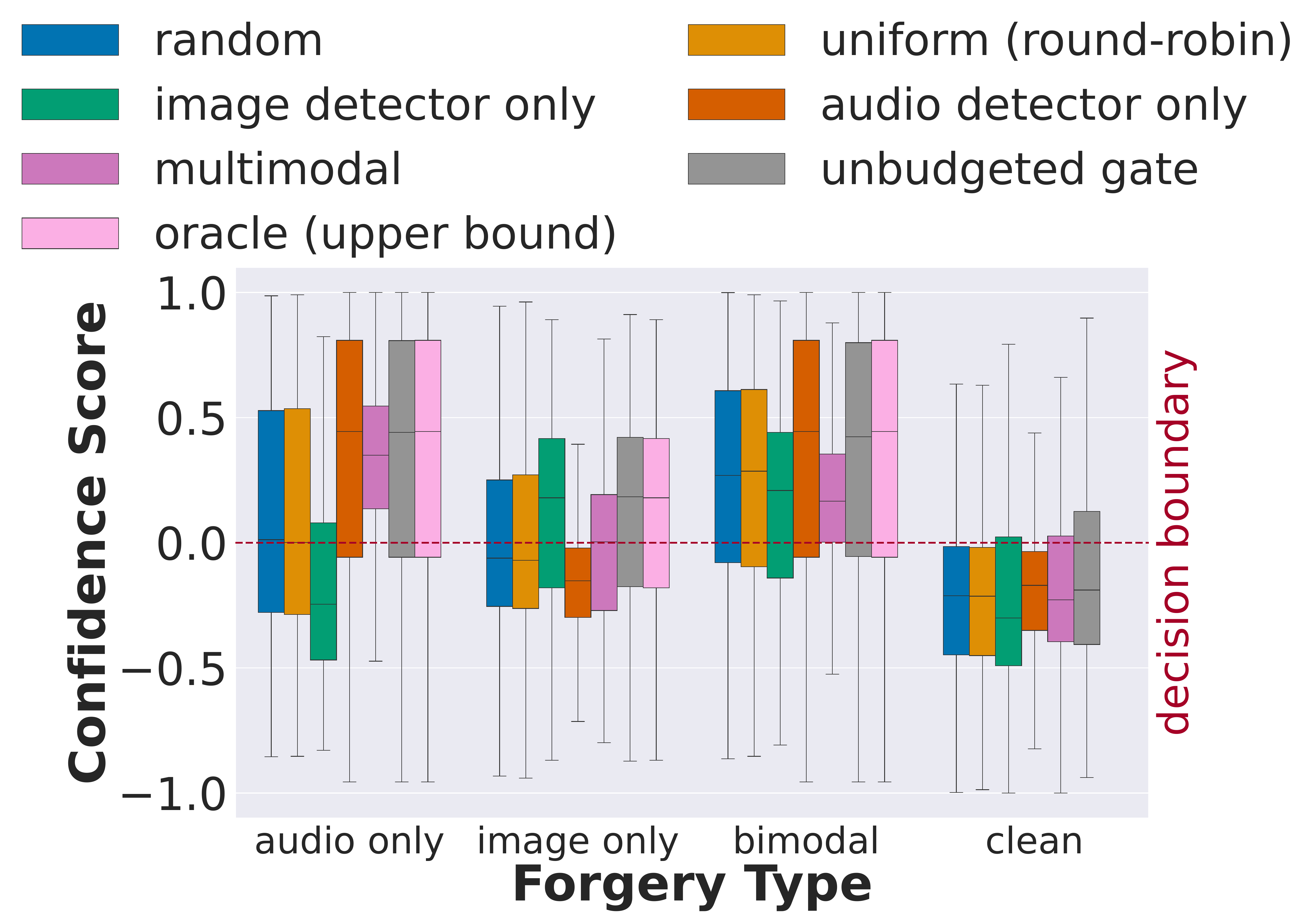}
  \caption{\textbf{With no budget cap, does picking the stream still matter?} Yes. The plot shows per-window confidence by forgery type, where confidence is rescaled to $[-1,1]$, so 0 marks the decision boundary. Every window is affordable, but at most one stream per window is read. The unbudgeted gate reads the stream that carries the edit. On audio edits, its median matches the audio detector at $+0.44$, while the image detector sits at $-0.25$. On image edits, it matches the image detector at $+0.18$, while the audio detector sits at $-0.15$. It reaches 0.819 accuracy, against 0.666, 0.635 and 0.718 for the audio, image and multimodal detectors alone, and 0.839 for the oracle.}
  \label{fig:margins}
\end{figure}

\vspace{2pt}\noindent\textbf{\underline{Predictions:}} A stream the router declines is never scored and defaults to authentic. Each acquired stream is thresholded at its detector's fixed operating point $\tau^m$, and the clip prediction max-pools the resulting margins:
\begin{equation}
\begin{aligned}
\hat y_t^m&=\mathds{1}[m\in S_t]\;\mathds{1}[s_t^m>\tau^m],\\
\hat y&=\mathds{1}\Big[\max_{t}\max_{m\in S_t}\big(s_t^m-\tau^m\big)>0\Big].
\end{aligned}
\label{eq:prediction}
\end{equation}
If nothing is acquired, $\hat y{=}0$. Pooling \textit{margins} rather than raw scores places detectors of unrelated scale on a common zero, and the \textit{maximum} lets a single forged window decide the clip. This is the opposite of averaging, which would dilute a two-second edit across a two-minute call. Nothing is lost in the pooling: every $\hat y_t^m{=}1$ is itself a finding that names \textit{when} and \textit{in which stream}.

Given $B$, we seek the router parameters $\theta_\varphi$ that maximize a reward $R$ that credits acquiring a manipulated stream, charges for acquiring an authentic one, and is neutral when nothing is acquired, so the budget acts as a ceiling rather than a target (expectation over videos and budgets):
\begin{equation}
\max_{\theta_\varphi}\ \mathbb{E}\Big[\sum_{t,m}R\big(\mathds{1}[m\in S_t],\,y_t^m\big)\Big]
\ \ \text{s.t.}\ \ 
\sum_{t=1}^{T}\sum_{m\in S_t} c^m \le B .
\label{eq:problem}
\end{equation}
The quota in Eq.~\ref{eq:problem} constrains the stream as a whole, yet each $S_t$ is committed before the remainder of the stream exists. Spending early risks having nothing left for the window that decides the outcome. Holding back risks never spending at all.

\vspace{2pt}\noindent\textbf{\underline{Threat model}} involves manipulating any subset of modalities over any subset of windows, at locations the adversary chooses. Edits are short and placed on the spans that carry meaning. The verifier knows neither how many edits a stream contains nor where they fall.

\vspace{2pt}\noindent\textbf{\underline{Assumptions:}} The detectors $\psi^m$ are pretrained, frozen, and treated as black boxes, so any scoring detector can be plugged in. The formulation admits any $c^m$. We charge one unit per detector call, $c^m{=}1$, so $B$ counts calls and a window read in both modalities costs two. The preview inside $\varphi$ is assumed cheap relative to any $\psi^m$. Per-window, per-stream labels are required for training only.

\vspace{2pt}\noindent\textbf{\underline{Decision loop:}} The router makes a single left-to-right pass with no lookahead as in Fig.~\ref{fig:arch}. At every window, it decides whether to spend and on which modality. Stopping needs no separate action. It is $S_t{=}\emptyset$, chosen freely or forced once $b_t$ reaches zero.

\vspace{2pt}\noindent\textbf{\underline{Preview and policy:}} The router $\varphi$ has two parts. A MobileNetV2 preview encoder~\cite{sandler2018mobilenetv2} reads a few subsampled frames and a spectrogram of each window and returns a compact feature, at a small fraction of a detector's cost. An LSTM~\cite{hochreiter1997lstm} cell carries what the stream has shown so far. At every step, it receives the new preview feature together with the remaining budget $b_t$ and horizon $T{-}t$, both as absolute counts, and a linear head scores \textit{four actions (none, audio, image, or both)}. The preview encoder is inherited from an unbudgeted per-window gate and kept frozen, so only the LSTM and the head, 2.37\,M parameters in total, are trained. Freezing the preview matters beyond cost. It fixes what the router can see, so every difference between the learned policies we compare comes from how they spend, not from what they perceive.

\vspace{2pt} The quota in Eq.~\ref{eq:problem} is enforced by construction rather than encouraged by a penalty. Before each decision, every action whose cost exceeds $b_t$ is masked by setting its logit to $-\infty$. With one unit left, \textit{both} disappears, and at $b_t{=}0$ only $\emptyset$ remains. The policy therefore chooses among affordable actions alone, and the quota holds for every stream and every budget, whether or not the policy is trained. A penalty would need a coefficient tuned to each budget and would only discourage violations. Masking needs no coefficient, adds no cost at inference, and applies unchanged during training, so the policy never learns from a trajectory the verifier could not afford.

\vspace{2pt}\noindent\textbf{\underline{Stage 1:}} Given the labels, Eq.~\ref{eq:problem} is a multiple-choice knapsack~\cite{sinha1979mckp}. Each window offers four actions, each with a cost and a reward, and exactly one is taken. It is solved exactly by dynamic programming~\cite{bellman1957dp} over (window, remaining budget), with $V(t,b)$ the best reward reachable from window $t$ with $b$ units left and $S\subseteq\mathcal{M}$,
\begin{equation}
\begin{aligned}
V(t,b)=\max_{S:\,\sum_{m\in S}c^m\le b}\Big[&\sum_{m\in\mathcal{M}}R\big(\mathds{1}[m\in S],\,y_t^m\big)\\
&+V\big(t{+}1,\;b-\textstyle\sum_{m\in S}c^m\big)\Big],
\end{aligned}
\label{eq:dp}
\end{equation}
with $V(T{+}1,\cdot)=0$, in $O(TB)$ time per stream. The table yields more than the optimal plan. It gives the optimal action from \textit{every} state $(t,b)$, including states that the optimal plan never visits. Stage~1 exploits this through DAgger distillation~\cite{ross2011dagger}. The student rolls out its own decisions under a sampled budget, and wherever it lands, mistakes included, the teacher supplies the best action from that exact state as a cross-entropy target. The student thus learns to recover from its own errors rather than to imitate a path it would never reproduce, and because budgets are sampled per stream, a single policy serves every budget. The teacher is clairvoyant. It reads $y_t^m$ for the whole stream, including windows the student has not reached, so part of what it knows cannot be learned from the preview alone.

\vspace{2pt}\noindent\textbf{\underline{Stage 2:}} Because the teacher's targets rest on information the student cannot have, stage~1 ends with a policy that imitates a foresight it lacks. Stage~2 therefore optimizes Eq.~\ref{eq:problem} directly, with self-critical policy gradient~\cite{williams1992reinforce,rennie2017scst} starting from the stage-1 weights. For each stream and budget, the policy samples one rollout and acts greedily in another. The greedy rollout's reward is the baseline, so a sampled decision is reinforced only when it beats what the policy would have done anyway. The baseline needs no learned critic and runs under the same mask and budget, so it is exactly as constrained as the sample it judges. Imitation thus supplies a competent start, and reinforcement adapts it to what the preview can actually see. In practice, stage~2 raises accuracy by 1.9 to 6.5 percentage points (pp) across budgets.

\section{Experimental Setup}

\noindent Here, we present our dataset, detectors, baselines, and metrics.\\
\noindent\textbf{\underline{Dataset:}} We use AV-Deepfake1M~\cite{cai2024avdeepfake1m}, chosen because its forgeries match our threat model. Unlike most earlier corpora~\cite{khalid2021fakeavceleb}, which forge whole clips, its manipulations are model-driven and content-driven: short segments of the audio, the image stream, or both are altered inside otherwise real videos, with per-stream time spans annotated. With over 1M videos of more than 2K subjects, it is large enough to train a router. Videos are divided into 1\,s frames at a 0.24\,s stride and split by identity. Results are reported on held-out videos the router never saw, in which 86.6\% of windows are authentic.

\begin{figure}[t]
  \centering
  \includegraphics[width=0.9\linewidth]{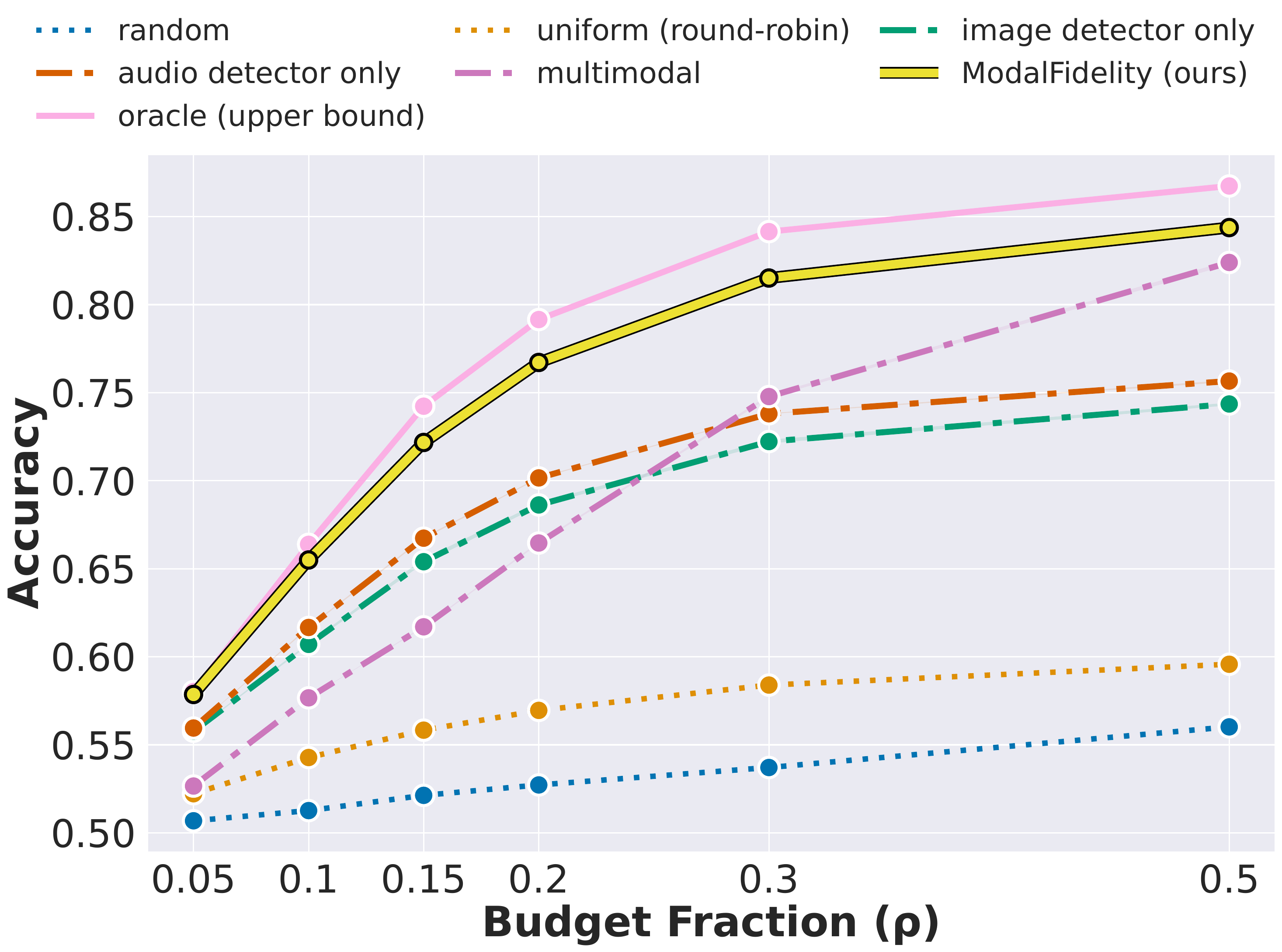}
  \caption{\textbf{Does spending more help without knowing where to look?} Accuracy against the budget fraction $\rho$. ModalFidelity stays within 2.6\,pp of the oracle at every budget and within 0.9\,pp for $\rho\le0.10$. At $\rho{=}0.10$ it reaches 0.655, against at most 0.617 for any fixed detector and 0.543 for the blind allocators, which stay near chance (0.51-0.60) however much they spend. Single-detector lines read one fixed stream on the router's windows.}
  \label{fig:curves}
\end{figure}

\vspace{2pt}\noindent\textbf{\underline{Detectors:}} The audio stream is scored by W2V2-AASIST~\cite{tak2022w2v2aasist,baevski2020wav2vec} and the image stream by GenD~\cite{yermakov2026gend} on a CLIP ViT-L backbone~\cite{radford2021clip}. AVH-Align~\cite{smeu2025avhalign}, which scores audio-visual consistency, serves as a multimodal alternative. All three are frozen and used as in Sec.~\ref{sec:method}.

\vspace{2pt}\noindent\textbf{\underline{Baselines}} receive the same budget and detectors as our router and differ only in how they spend. The \textit{oracle} is the teacher of Eq.~\ref{eq:dp} run on test labels, a ceiling no causal policy can reach. \textit{Audio-only}, \textit{image-only} and \textit{multimodal} keep our router's windows but force a single detector on each. \textit{Uniform} and \textit{random} spend blindly. We also compare against the unbudgeted gate our preview is inherited from, and against late MoE gates over both detectors' embeddings, which run both detectors on every window.

\vspace{2pt}\noindent\textbf{\underline{Metrics:}} We score the per-window predictions of Eq.~\ref{eq:prediction} against the annotated spans and report detection accuracy, the mean of recall on manipulated windows and specificity on authentic ones, averaged over windows, or per video where stated. Because every window and stream is scored against the annotated spans, this accuracy measures localization at window resolution. Cost is the budget fraction $\rho$ and, when comparing gate placements, GFLOPs per window.
\section{Results}
\label{sec:results}

\noindent Here, we answer five questions about where the evidence lies, what a budget buys, and where the gate belongs.\\

\noindent\textit{\underline{Which stream carries the evidence?}}

\noindent Fig.~\ref{fig:margins} groups every window by which stream was forged, and counting how often each detector fires shows the same asymmetry. The audio detector fires on 70\% of audio edits but on only 22\% of image ones, no more often than its 21\% rate on authentic windows. The image detector mirrors it, firing on 64\% of image edits and 30\% of audio ones, against a 26\% rate on authentic windows. Each detector is thus effectively blind to the stream it does not read. Even AVH-Align, which reads both, catches 88\% of audio edits but only half of the image ones. No fixed choice of detector covers both kinds of edit, so which stream to read must be decided window by window, which is exactly the decision the router makes.\\

\noindent\textit{\underline{Does more budget buy more accuracy?}} 

\noindent Our router stays within 2.6\,pp of the clairvoyant oracle at every budget as in Fig.~\ref{fig:curves}. At $\rho{=}0.30$ it reaches an accuracy of 0.815, within 0.4\,pp of the 0.819 reached by the unbudgeted gate, which is bound by no budget at all. At $\rho{=}0.50$, where it may read both streams of a window, it passes the gate with 0.844. Blind allocators show the opposite. Uniform and random stay near chance, between 0.51 and 0.60, and gain little as the budget grows, because most windows are authentic and a blind acquisition rarely lands on an edit. Budget is an asset only for a policy that knows where to spend it.\\

\noindent\textit{\underline{Which strategy wins where, one video at a time?}}

\begin{figure}[t]
  \centering
  \includegraphics[width=0.9\linewidth]{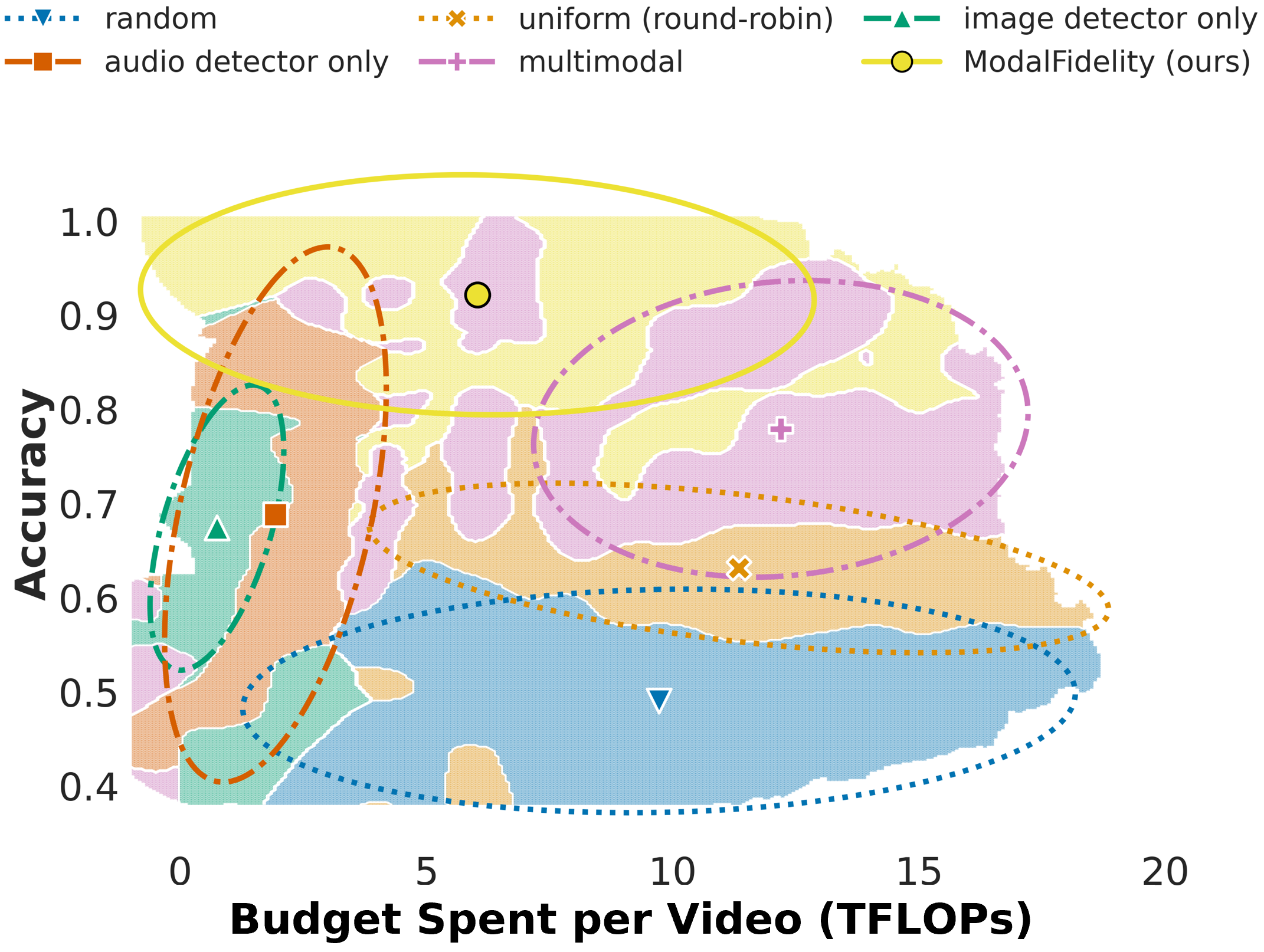}
  \caption{\textbf{Which strategy wins where, one video at a time?} We measure the performance of each method on a per-video basis to obtain the pairs (cost, accuracy). Shading marks the method most common in that pair, and each ellipse traces that method's largest area. The blind allocators own the accuracy below ${\approx}0.65$, the single detectors the low-spend strip under 4 TFLOPs, multimodal the middle band, and ModalFidelity the high-accuracy area with low compute cost.}
  \label{fig:knn_ellipses}
\end{figure}

\noindent Fig.~\ref{fig:knn_ellipses} places every forged video by cost and accuracy: the blind allocators own the band below 0.65, the single detectors the low-spend strip, and our router's area sits at the top, beside the oracle's. At $\rho{=}0.20$, uniform and random exhaust their budgets on every forged video, 7.24 units on average, while our router spends 6.27 units on average, or 86\% of what it is allowed, and still reaches 0.771 per-video accuracy against their 0.577 and 0.526. Only the clairvoyant oracle is more accurate, at 0.799, with nearly the same spend. The single-detector baselines spend less only because each of our two-stream windows costs them one unit, and they lose 6 to 9\,pp per video for it. At clip level, flagging a video when any of its windows is flagged, our router reaches 0.928 at $\rho{=}0.20$, against 0.940 for the oracle and at most 0.875 for any fixed detector. \textit{A real stream costs budget and returns nothing, so the router is free to hold back, treating the budget as a ceiling rather than a target.}\\

\noindent\underline{\textit{What does a fixed budget buy elsewhere?}} 

\noindent Fig.~\ref{fig:curves} holds the budget fixed and varies only how it is spent across windows and streams. The single-detector baselines keep our router's windows but always read the same stream. They trail our router at every budget, by 6.6 to 8.1\,pp at $\rho{=}0.20$, so choosing the stream adds accuracy on top of choosing the window.\\

% [commented out 2026-09-23, user: use the k-NN + ellipse figure only]
% \begin{figure}[t]
%   \centering
%   \includegraphics[width=\linewidth]{fig4_regions.pdf}
%   \caption{\textbf{Per-video cost and accuracy at $\rho{=}0.20$.} Each region covers about half of a method's forged videos; markers are means, and up and to the left is better. Our router sits just below the oracle while leaving part of its budget unspent; the blind allocators spend all of theirs.}
%   \label{fig:regions}
% \end{figure}

% [commented out 2026-09-23, user: use the k-NN + ellipse figure only]
% \begin{figure}[t]
%   \centering
%   \includegraphics[width=\linewidth]{fig5_knn_regions.pdf}
%   \caption{\textbf{Which method does a video most likely come from?} The same videos as Fig.~\ref{fig:regions}, with each point of the plane shaded by the method whose videos dominate its neighbourhood ($k$-NN, $k{=}150$); markers are means. Accuracy here is balanced.}
%   \label{fig:knn}
% \end{figure}

% [commented out 2026-09-23, user: keep only k=150]
% \begin{figure}[t]
%   \centering
%   \includegraphics[width=\linewidth]{fig5_knn_regions_k5.pdf}
%   \caption{\textbf{As Fig.~\ref{fig:knn}, with $k{=}5$.}}
%   \label{fig:knn_k5}
% \end{figure}

% [commented out 2026-09-23, user: keep only k=150]
% \begin{figure}[t]
%   \centering
%   \includegraphics[width=\linewidth]{fig5_knn_regions_k500.pdf}
%   \caption{\textbf{As Fig.~\ref{fig:knn}, with $k{=}500$.}}
%   \label{fig:knn_k500}
% \end{figure}

% [commented out 2026-09-23, user: keep only k=150]
% \begin{figure}[t]
%   \centering
%   \includegraphics[width=\linewidth]{fig5_knn_regions_k1000.pdf}
%   \caption{\textbf{As Fig.~\ref{fig:knn}, with $k{=}1000$.}}
%   \label{fig:knn_k1000}
% \end{figure}

\begin{table}[t]
  \caption{\textbf{Why route before the detectors rather than after?}}
  \label{tab:placement}
  \centering\small\setlength{\tabcolsep}{3pt}
  \begin{tabular}{llccc}
    \toprule
    Placement & & Calls/win.\,($\downarrow$) & GFLOPs/win.\,($\downarrow$) & Acc.\,($\uparrow$) \\
    \midrule
    Late MoE & dense   & 2.000 & 4086.6 & .7264 \\
          & feature & 2.000 & 4086.6 & .7298 \\
          & output  & 2.000 & 4086.6 & .7322 \\
    \midrule
    Input (ours) & $\rho{=}0.20$ & 0.130 & \phantom{0}257.5 & .7672 \\
                 & $\rho{=}0.30$ & 0.161 & \phantom{0}327.3 & .8151 \\
                 & $\rho{=}0.50$ & 0.183 & \phantom{0}384.5 & \textbf{.8444} \\
    \bottomrule
  \end{tabular}
\end{table}

\noindent\textit{\underline{What does moving the gate save?}} 

\noindent In Table~\ref{tab:placement}, each late gate runs as designed, with both detectors on every window, which costs 4,087 GFLOPs per window wherever the gate sits. At the input, our router needs 258 GFLOPs at $\rho{=}0.20$, \textbf{15.9$\times$} fewer, and 15.4$\times$ fewer detector calls, while its accuracy is higher, at 0.767 against at most 0.732. At $\rho{=}0.50$ it is 10.6$\times$ cheaper and 11\,pp more accurate. Among the late gates, position moves accuracy by under 0.6\,pp. We compare placements, not training recipes: the late heads are supervised classifiers, and the router is trained to acquire. A gate pays for itself before the detectors run.

\section{Conclusion}

In this paper, we asked how a verifier can expose targeted deepfakes by verifying only where it matters, and answered with \textbf{ModalFidelity}, a lightweight router that decides window by window which stream to acquire under a hard compute budget. On AV-Deepfake1M the router retains over \textbf{96\%} of the accuracy of a clairvoyant oracle at every budget and needs \textbf{15.9$\times$} fewer FLOPs than late MoEs. Its output is not a single label but a per-window, per-stream prediction that states, at window resolution, \textit{when} a forgery occurred and \textit{in which stream}. We expect the principle to carry beyond deepfakes, wherever evidence is sparse and costly to read: deciding where to look is worth more than looking harder. Future work involves an adversary that learns the routing policy and targets the windows it skips. Our code, models, and benchmarks are available$^\text{\ref{code}}$.

\bibliographystyle{IEEEbib}
\bibliography{ref/swarm,ref/refs}

\end{document}